\documentclass[a4paper,11pt]{article}
\pdfoutput=1 
 
\usepackage{jcappub} 
 
\usepackage{aas_macros}
\usepackage{amssymb,amsmath}
\usepackage{graphicx}
\usepackage[dvipsnames]{xcolor}
\hypersetup{colorlinks=true,allcolors=teal}

\newcommand{\be}{\begin{equation}}
\newcommand{\ee}{\end{equation}}
\def\bear#1\ear{\begin{align}#1\end{align}}
\newcommand{\nline}{\notag \\}
\newcommand{\f}{\frac}
\newcommand{\de}{\mathrm{d}}

\renewcommand{\mathbf}[1]{\mbox{\boldmath $#1$}}

\newcommand{\Msun}{\mathrm{M}_{\odot}}
\newcommand{\hcMpc}{h^{-1} \mathrm{cMpc}}

\newcommand{\eqn}[1]{equation~(\ref{#1})}

\newcommand{\secn}[1]{Section~\ref{#1}}
\newcommand{\fig}[1]{Figure~\ref{#1}}
\newcommand{\figs}[2]{Figures~\ref{#1} and \ref{#2}}
\newcommand{\figsthree}[3]{Figures~\ref{#1}, \ref{#2} and \ref{#3}}

\defcitealias{2023MNRAS.526.3920M}{MPC23}

\renewcommand{\citet}{\cite}

\title{A GPR-Based Emulator for Semi-numerical Reionization Code SCRIPT: Parameter Inference from 21~cm Data}

\author[a,1]{T. Roy Choudhury,\note{Corresponding author.}}
\author[b]{A. Paranjape,}
\author[a]{and B. Maity}
 
\affiliation[a]{National Centre for Radio Astrophysics, TIFR,\\Pune University Campus, Ganeshkhind, Pune 411007, India}
\affiliation[b]{Inter-University Centre for Astronomy \& Astrophysics,\\Pune University Campus, Ganeshkhind, Pune 411007, India}
 
\emailAdd{tirth@ncra.tifr.res.in}
\emailAdd{aseem@iucaa.in}
\emailAdd{barun.m003@gmail.com}

\abstract{
Semi-numerical models of reionization typically involve a large number of unknown parameters whose values are constrained by comparing with observations. Increasingly often, exploring this parameter space using semi-numerical simulations can become computationally intensive, thus necessitating the use of emulators. In this work, we present a likelihood emulator based on Gaussian Process Regression (GPR) for our semi-numerical reionization code, \texttt{SCRIPT}, and use it for parameter inference using mock 21~cm power spectrum data and Bayesian MCMC analysis. A unique aspect of our methodology is the utilization of coarse resolution simulations to identify high-probability regions within the parameter space, employing only a moderate amount of computational time. Samples drawn from these high-probability regions are used to construct the training set for the emulator. The subsequent MCMC using this GPR-trained emulator is found to provide parameter posteriors that agree reasonably well with those obtained using conventional MCMC. The computing time for the analysis, which includes both generation of training sets and training the emulator, is reduced by approximately an order of magnitude. This methodology is particularly advantageous in scenarios where one wants to use different parametrizations of reionization models and/or needs to start with broad prior distributions on the parameters, offering an efficient and effective means of parameter inference.
}


\begin{document} 
 
\date{} 

\maketitle

\flushbottom

\section{Introduction}

Semi-numerical simulations of the epoch of reionization have proven to be extremely effective in exploring the space of unknown model parameters related to the high redshift universe. A promising way of constraining these parameters is by comparing the 21~cm fluctuation signals with existing and upcoming data, making use of Bayesian parameter estimation techniques \cite{2015MNRAS.449.4246G,2016MNRAS.455.4295G,2017MNRAS.472.2651G,2018MNRAS.477.3217G,2019MNRAS.484..933P,2021MNRAS.500.5322G,2021MNRAS.501....1G,2023MNRAS.521.4140M}. However, sometimes these models involve a large number of parameters, and that can lead to extended computation times, especially when the simulations are of high dynamic range.

With recent advancements in the field of machine learning, several techniques have been proposed to deal with such challenges. One option is to use an emulator for the simulations that can compute the 21~cm observables, or any other quantity relevant for the parameter inference, without running the actual simulation \cite{2017ApJ...848...23K,2017MNRAS.468.3869S,2018MNRAS.475.1213S,2019MNRAS.483.2907J,2020MNRAS.493.4728G,2020MNRAS.498.4178M,2022MNRAS.514L..31M,2023arXiv230715577L,2023MNRAS.524.4239P,2023arXiv230905697B}. The approach requires running the simulation at some selected set of points in the parameter space, known as the training set. Subsequently, one can train the emulator so that it can predict the quantities at any other point in the parameter space. Developing emulators for expensive simulations is of great use for exploring parameter space, particularly when the problem has a high dimensionality. Almost all the studies available in the literature have discussed the emulation of the observable (e.g., the 21~cm power spectrum) for use in parameter inference. In this study, we explore the potential of a likelihood emulator developed for our explicitly photon-conserving semi-numerical reionization model \texttt{SCRIPT} \cite{2018MNRAS.481.3821C}, in the context of upcoming 21~cm observations. In our previous work \citet[][henceforth, MPC23]{2023MNRAS.526.3920M}, we introduced an emulator based on Gaussian Process Regression (GPR) mainly for parameter inference with existing observational data. This work is a follow up where we assess the performance of the emulator with mock 21~cm data, appropriate for upcoming telescopes like the SKA\footnote{\url{https://www.skao.int/index.php/en/science-users}}.

We showed in \citetalias{2023MNRAS.526.3920M} that the resources required for the emulator-based parameter inference can be reduced significantly if one selects the training set carefully, ideally in regions of high probability within the parameter space. We proposed utilizing simulations that require significantly less computing resources, such as those of coarser resolutions compared to the default, to run a preliminary parameter estimation, and thus obtain an approximate posterior. This distribution can serve as a basis for selecting the training samples. Our results indicate that this method provides excellent match with parameter posteriors obtained using a traditional analysis, but requiring significantly less computing time.

In this project, we repeat the analysis but for 21~cm data relevant for the SKA. We generate mock data using our model, incorporating telescope noise and foreground avoidance techniques. Subsequently, we apply the GPR-based emulator to obtain the parameter posteriors from the mock data using a Bayesian Markov Chain Monte Carlo (MCMC) method. The resulting posteriors are then compared with the ones obtained using a full MCMC without any reference to the emulator. This comparison also allows us to understand the gain in computing time while using the emulator.

An interesting feature of our method is that one can build the emulator from scratch for arbitrary parametrizations of the model; one does not need any prior knowledge of the parameters or access to any pre-existing MCMC chains. The only requirement for our analysis is that the simulation results should be converged with respect to the resolution of the simulation, i.e., the coarse resolution less resource-intensive simulations should not lead to any significant bias in the parameter posteriors compared to the default runs.

The paper is organized as follows: In \secn{sec:method}, we highlight the essential components of our semi-numerical model \texttt{SCRIPT}, and use it to generate the mock 21~cm power spectra data. Furthermore, present parameter inferences using conventional MCMC methods, which will serve as benchmarks for comparing the results obtained using the emulator. In \secn{sec:GPR}, we discuss in detail the GPR-based emulator developed for \texttt{SCRIPT}, along with a comparative analysis against the full MCMC results. To conclude, in \secn{sec:summary}, we summarize our findings and discuss potential avenues for future extensions.
The cosmological parameters used in this work are $\Omega_m = 0.308, \Omega_{\Lambda} = 1- \Omega_m, \Omega_b = 0.0482, h = 0.678, n_s = 0.961, \sigma_8 = 0.829$ \citep{2016A&A...594A..13P}.

\section{Method}
\label{sec:method}

\subsection{Generation of 21~cm light cones using \texttt{SCRIPT}}
\label{subsec:ionization_maps}

The theoretical model of reionization used in this work is based on the semi-numerical code, named \textbf{S}emi-numerical \textbf{C}ode for \textbf{R}e\textbf{I}onization with \textbf{P}ho\textbf{T}on-conservation (\texttt{SCRIPT})\footnote{\url{https://bitbucket.org/rctirthankar/script/}} \cite{2018MNRAS.481.3821C}. The method requires two sets of inputs at all relevant redshifts $z$:

\begin{enumerate}

\item The density $\Delta_m(\mathbf{x}, z)$ and velocity $\mathbf{v}(\mathbf{x}, z)$ fields of the matter smoothed over a grid, with the cells labelled by their comoving coordinates $\mathbf{x}$, where $\Delta_m(\mathbf{x}, z) \equiv \rho_m(\mathbf{x}, z) / \bar{\rho}_m$ is the ratio of matter density to the corresponding mean.

\item The mass and spatial distribution of haloes that can host star-forming galaxies.

\end{enumerate}

In this work, we generate the $\Delta_m$ and $\mathbf{v}$ fields using a 2LPT code \texttt{MUSIC}\footnote{\url{https://www-n.oca.eu/ohahn/MUSIC/}} \cite{2011MNRAS.415.2101H}, assuming all matter to be collisionless. This allows us to generate redshift snapshots with minimal computational cost, and we have found that the fields obtained agree better than $\sim 2\%$ with the full $N$-body simulations for grid sizes $\gtrsim 1 \hcMpc$, sufficient for our purpose.

The main disadvantage of using simulations that are of relatively coarser resolutions is that the small mass haloes $M \lesssim 10^8 \Msun$, that are believed to be the hosts of the first stars, cannot be identified. We hence use a sub-grid formalism to populate the grid cells with haloes. The formalism is based on conditional ellipsoidal collapse model \cite{2002MNRAS.329...61S}, and provides reasonably accurate halo mass functions and large-scale halo clustering; for details we refer the reader to our earlier work \cite{2018MNRAS.481.3821C}. This approximate way of computing the collapsed mass provides reasonable match with the full simulations only for relatively larger grid volumes, hence we do not use grids finer than $2 \hcMpc$.

The generation of ionization maps at a given redshift requires the ionization efficiency $\zeta(M, z)$, which is defined as the number of ionizing photons in the IGM per hydrogen in stars, and can be a function of both halo mass $M$ and redshift $z$. The algorithm implemented in \texttt{SCRIPT} conserves the number of photons explicitly and proceeds in two steps: in the first we generate ionized regions around sources, allowing regions that receive photons from multiple sources to have an ionization fraction greater than unity. In the second step, we redistribute the excess photons in these over-ionized cells to nearby regions, thus ensuring photon conservation. The advantage of this algorithm is that it provides large-scale power spectra that are converged with respect to the size of the grid cell \cite{2018MNRAS.481.3821C}.

Once the ionization map is generated, we have access to the neutral fraction $x_\mathrm{HI}(\mathbf{x}, z)$ on a uniform grid, in addition to $\Delta_m$ and $\mathbf{v}$ obtained from the 2LPT code. The 21~cm brightness temperature at every grid cell, accounting for the redshift space distortions, can be obtained as \cite{2004MNRAS.352..142B,2006PhR...433..181F}
\be
\delta T_b(\mathbf{x}, z) = 27 \text{mK}~x_\mathrm{HI}(\mathbf{x}, z) \Delta_m(\mathbf{x}, z) \left[1 + \f{1}{H(z)}\f{\de v_r(\mathbf{x}, z)}{\de r} \right]^{-1}
\left(\f{1 + z}{10} \f{0.15}{\Omega_m h^2}\right)^{1/2} \left(\f{\Omega_b h^2}{0.023}\right),
\ee
where $v_r$ is the line of sight component of the velocity $\mathbf{v}$ and $\de v_r / \de r$ is the gradient with respect to the comoving distance along the line of sight. We have assumed that the spin temperature of the gas is significantly higher than the background CMB temperature, appropriate for the epoch of reionization.

We generate ionization maps from $z = 20$ to $z = 5$ with 151 snapshots equally spaced in the scale factor $a$ with an interval $\Delta a \approx 8 \times 10^{-4}$. Having snapshots equi-spaced in $a$ allows us to finely sample the redshifts where the ionization fraction increases sharply for a typical reionization history. However the conclusions of the paper are not sensitive to the sampling of redshifts as long as $\Delta z \lesssim 0.1$. Our simulation boxes are of length $128 \hcMpc$, sufficient to probe the smallest $k$-modes $\sim 0.2~h$\,cMpc$^{-1}$ used in this work. Our default simulations have a grid size $2 \hcMpc$, hence consists of $64^3$ grids. We will refer to these as ``default resolution simulations''. We also use simulations of coarse resolution having only $32^3$ grids with a grid size of $4 \hcMpc$.

We parametrize the ionization efficiency as
\bear
\zeta(M, z) &= \zeta_0 \left(\f{1 + z}{8}\right)^{\alpha_{\zeta, z}} && \text{for } M > M_\mathrm{min}(z),
\nline
&= 0 && \text{otherwise}
\ear
where the minimum mass of haloes that can host star-forming galaxies is assumed to be of the form
\be
M_\mathrm{min}(z) = M_{\mathrm{min}, 0} \left(\f{1 + z}{8}\right)^{\alpha_{M_\mathrm{min}, z}}.
\ee
Our model thus has four free parameters, namely, $\zeta_0, \alpha_{\zeta, z}, M_{\mathrm{min}, 0}, \alpha_{M_\mathrm{min}, z}$.

Note that the efficiency parameter is independent of $M$ for $M > M_\mathrm{min}(z)$, as has a step-like cutoff at $M =  M_\mathrm{min}(z)$. It is possible to accommodate more complicated $M$-dependence \cite{2013MNRAS.432.3340S,2019MNRAS.482L..19C,2021MNRAS.506.2390Q}, however, we find that the corresponding free parameters cannot be constrained using only 21~cm data. In general, constraining the $M$-dependence requires using data sets that respond to the halo mass more directly, e.g., galaxy luminosity function, which we do not consider in this work. Similarly, we do not include other physical effects like inhomogeneous recombinations and radiative feedback \cite{2022MNRAS.511.2239M,2022MNRAS.515..617M} because the 21~cm data used in this work cannot provide any meaningful constraints on the corresponding free parameters.

At each redshift snapshot, we generate the ionization map for given values of $\zeta(z)$ and $M_\mathrm{min}(z)$. We then generate a light cone extending from $z = 20$ to $z = 5$ by interpolating the fields obtained for the comoving boxes. However, since our model does not contain the spin temperature fluctuations, it is not suitable for very early stages of reionization. In general, the effect of X-ray heating fluctuations become relatively less significant when the global ionized fraction $\gtrsim 10\%$ \cite{2015MNRAS.447.1806G}. For the typical reionization histories considered in this work, we find that this threshold fraction is reached at $z \lesssim 11.5$, hence this serves as the upper limit of the redshift range used in our analysis. At the other end, the analysis is restricted to $z \gtrsim 5.5$ as there is almost no 21~cm signal at lower redshifts. The light cone used for subsequent calculations has a comoving length of $\sim 1.15 h^{-1}$cGpc.

Since our comoving box is only of size $128 \hcMpc$, we need to find a way to extend the fields beyond the box size while constructing the light cone. Repeating the periodic box preserves the continuity in the underlying fields, however, this method may lead to repeating structures and consequently give rise to spurious correlation signals. We hence use a different realization of the density field (i.e., fields generated using a different initial condition of the cosmological fields) for joining the boxes once the length along the line of sight extends beyond $128 \hcMpc$. The ionization field thus constructed is discontinuous at the border of the two boxes, and the correlation between the fields are not accounted for beyond the box length. These, however, are unlikely to cause any serious concerns as we will see below.

We next divide the light cone into frequency bands, which can subsequently be used for computing the summary statistics like the 21~cm power spectrum. An optimal value of the band width has been found to be $\Delta \nu \sim 10$~MHz \cite{2014MNRAS.442.1491D}. Such values are small enough that the signal does not evolve significantly within the band, at the same time providing enough volume to probe the largest scales of interest and also reduce the telescope noise. We find that our box length corresponds to $\Delta \nu \approx 13.25$~MHz at $z \sim 6$ and $\approx 9.79$~MHz at $z \sim 11$. These are quite similar to the optimal values. We thus divide the light cone into equal comoving lengths of $128~h^{-1}$~cMpc leaving us with convenient cubical boxes to deal with. Each of these volumes is continuous, with the discontinuity occurring only at the boundaries of the two different frequency bands. This approach also allows us to ignore the discontinuities in the fields at the boundaries of the different boxes. We end up with 9 frequency bands at our disposal for the analysis, each labelled by the redshift range $[z_\mathrm{min}, z_\mathrm{max}]$ covered.

Because the realization of the density field is different across different frequency bands, the covariance of the power spectra across different bands will not be accounted for. Note that in order to account for such covariances properly, we would ideally need a box of length $\gtrsim 1 h^{-1}$~cGpc, and that would make the parameter exploration extremely resource-intensive. It has been found that, after adding the observational effects like the telescope noise and foreground avoidance, the covariance terms between different frequency bands are negligible, except adjacent bins at the lowest redshifts $z \sim 6$ \cite{2023MNRAS.524.4239P}. In this work, we neglect these covariance terms and work with the assumption that different frequency bands are uncorrelated. See, however \citet{2023arXiv231016464A} for an assessment of the potential impact of such effects in a different context.

For generating the light cone and subsequently the mock power spectra (in \secn{subsec:mock_data}), we need to choose the fiducial values of the free parameters of our model. In this work, we choose $\zeta_0 = 8$, $\alpha_{\zeta, z} = 0$, $M_{\mathrm{min}, 0} = 10^8 \Msun$ and $\alpha_{\mathrm{min}, z} = 0$. The resulting reionization history leads to a CMB optical depth $\tau = 0.058$, consistent with the existing observational constraints \cite{2016A&A...596A.108P}. Reionization is completed at $z \approx 5.6$ in this case, consistent with observations of quasar absorption spectra that require late completion of reionization \cite{2018ApJ...864...53E,2018MNRAS.479.1055B,2019ApJ...881...23E,2021ApJ...923...87C,2021ApJ...923..223Z,2022MNRAS.514...55B}. Our fiducial choice of parameters correspond to redshift-independent $\zeta$ and $M_\mathrm{min}$. In some of our previous studies \cite{2021MNRAS.501L...7C,2022MNRAS.515..617M,2023MNRAS.522.2901J}, we found that current observations prefer a $\zeta$ that increases with redshift, although we could not rule out the possibility of non-evolving $\zeta$. Similarly, the minimum mass $M_\mathrm{min}$ is also expected to evolve because of feedback and other effects, however, its exact evolution is not fully understood.  In the absence of any comprehensive understanding of the temporal evolution of these quantities, our assumption of their constancy can be thought of the simplest and most straightforward approach. It is worth noting that our conclusions regarding the effectiveness of the GPR-based emulator are independent of our choice of the fiducial parameters.

\begin{figure*}
\centering
\includegraphics[width=\textwidth]{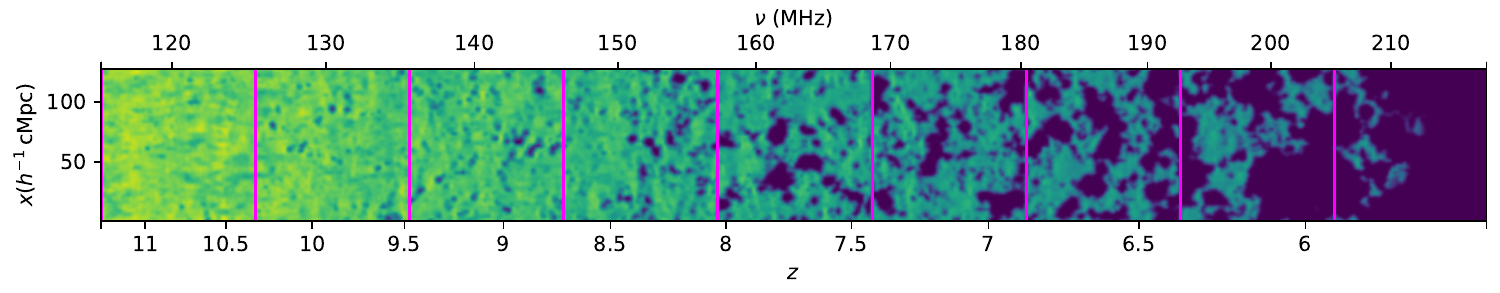}
\caption{A slice through the 21~cm light cone generated for the fiducial reionization history. Green (dark) regions correspond to higher (lower) values of the differential brightness temperature $\delta T_b$. The vertical magenta lines denote the frequency bands used for computing the summary statistics.}
\label{fig:lightcone}
\end{figure*}

A representative slice through the light cone for the fiducial model generated at the default grid size is shown in \fig{fig:lightcone}.

\subsection{Observational effects and generation of mock 21~cm data}
\label{subsec:mock_data}

Having generated the light cone for the brightness temperature, we add observational effects to the simulation for further analysis. In the first step, we add telescope noise as appropriate for the upcoming SKA-Low\footnote{\url{https://www.skao.int/index.php/en/science-users/118/ska-telescope-specifications}}. We assume a tracked scan at a declination of $-30^{\circ}$ of 6 hours per day for 180 days, thus giving a total integration time of 1080 hours. The noise is computed using the publicly available \texttt{tools21cm} package\footnote{\url{https://github.com/sambit-giri/tools21cm}} \cite{2017MNRAS.464.2234G,2018MNRAS.479.5596G}. We subtract the mean from each frequency slice to realistically simulate the absence of zero baseline modes. The power spectrum $P_{21}(\mathbf{k}; z_\mathrm{min}, z_\mathrm{max})$ of this mean subtracted $\delta T_b$ is calculated for each frequency band labelled by the redshift range $[z_\mathrm{min}, z_\mathrm{max}]$. Note that the power spectrum is anisotropic because of line of sight effects like the redshift space distortion and evolution along the light cone.

The quantity of our interest is the spherically averaged power spectrum, computed by averaging $P_{21}(\mathbf{k}; z_\mathrm{min}, z_\mathrm{max})$ over angles
\be
P_{21}(k; z_\mathrm{min}, z_\mathrm{max}) = \int \f{\de \Omega}{4 \pi} P_{21}(\mathbf{k}; z_\mathrm{min}, z_\mathrm{max})
= \int_0^1 \de \mu~P_{21}(\mathbf{k}; z_\mathrm{min}, z_\mathrm{max}),
\ee
where $\mu \equiv k_\parallel / k$ is the cosine of the angle between the wave vector $\mathbf{k}$ and the line of sight, $k_\parallel$ being the $\mathbf{k}$-component along the line of sight. A more relevant quantity is the power per logarithmic $k$-interval
\be
\Delta_{21}^2(k; z_\mathrm{min}, z_\mathrm{max}) = \f{k^3 P_{21}(k; z_\mathrm{min}, z_\mathrm{max})}{2 \pi^2},
\ee
which has a dimension same as $\delta T_b^2$.

In order to account for the foregrounds, we ignore the all modes below the horizon limit in the ``foreground wedge'' \cite{2010ApJ...724..526D,2012ApJ...745..176V,2012ApJ...752..137M,2012ApJ...757..101T,2012ApJ...756..165P,2013ApJ...768L..36P,2013ApJ...770..156H,2014PhRvD..90b3018L,2014PhRvD..90b3019L,2015ApJ...804...14T,2018ApJ...869...25M,2020PASP..132f2001L}
\be
k_{\parallel} \leq k_{\perp} C, \quad C = \f{x(z) H(z)}{c (1 + z)},
\ee
where $k_\perp$ is the component of $\mathbf{k}$ perpendicular to the line of sight and $x(z)$ is the comoving distance to redshift $z$. The above condition can be written in terms of $\mu$ as
\be
\mu \leq \mu_\mathrm{min} \equiv \f{C}{\sqrt{1 + C^2}},
\ee
and hence ignoring the corresponding modes leads to a modified expression for the spherically averaged power spectrum
\be
P_{21}(k; z_\mathrm{min}, z_\mathrm{max}) = \f{1}{1 - \mu_\mathrm{min}} \int_{\mu_\mathrm{min}}^1 \de \mu~P_{21}(\mathbf{k}; z_\mathrm{min}, z_\mathrm{max}).
\ee
Note that since modes with $\mu \leq \mu_\mathrm{min}$ are \emph{not} included while calculating $P_{21}(k; z_\mathrm{min}, z_\mathrm{max})$ and $\Delta^2_{21}(k; z_\mathrm{min}, z_\mathrm{max})$, the resulting power spectra are expected to be biased compared to the true ones, particularly at small-$k$ values \citep{2016MNRAS.456...66J,2018MNRAS.475..438R}. 

To compute the mock 21~cm data for the nine frequency bands, we choose 9 random realizations of the simulation and assign each realization to a specific frequency band. The corresponding $\Delta_{21}^2(k; z_\mathrm{min}, z_\mathrm{max})$ for the different frequency bands constitute the mock data. For the default resolution simulations, we bin the power spectra into 7 bins equi-spaced in $\log_{10} k$, with bin centres at
\be
\log_{10} \left(\f{k_i}{h\,\text{cMpc}^{-1}}\right) = 0.65 + 0.1~(i - 1), \quad i = 1, \ldots, 7.
\ee
This binning allows us to probe $k$-values between $0.2 h\,\text{cMpc}^{-1}$ and $1 h\,\text{cMpc}^{-1}$, scales that can be reliably probed by upcoming observations in the presence of telescope noise and foregrounds. The total number of data points $N_\mathrm{data}$ across all frequency bands is thus 63. For the coarse resolution simulations, we ignore the two largest $k$-bins and so are left with only 5 data points per frequency band, i.e., $N_\mathrm{data} = 45$.

The observational errors on $\Delta^2_{21}(k_i; z_\mathrm{min}, z_\mathrm{max})$ are computed by generating $50$ realizations of the data (with different initial conditions of the cosmological fields and different seeds for the telescope noise). The corresponding covariance matrix $\Sigma^\mathrm{data}_{ij}$, where $i,j = 1, 2, \ldots, N_\mathrm{data}$, provides the quantity used for likelihood calculation. Note that $\Sigma^\mathrm{data}_{ij}$ contains the contribution of both telescope noise and cosmic variance. As discussed earlier, we ignore the covariance between different frequency bands, thus $\Sigma^\mathrm{data}_{ij}$ in our case has a block-diagonal form.

\begin{figure*}
\centering
\includegraphics[width=\textwidth]{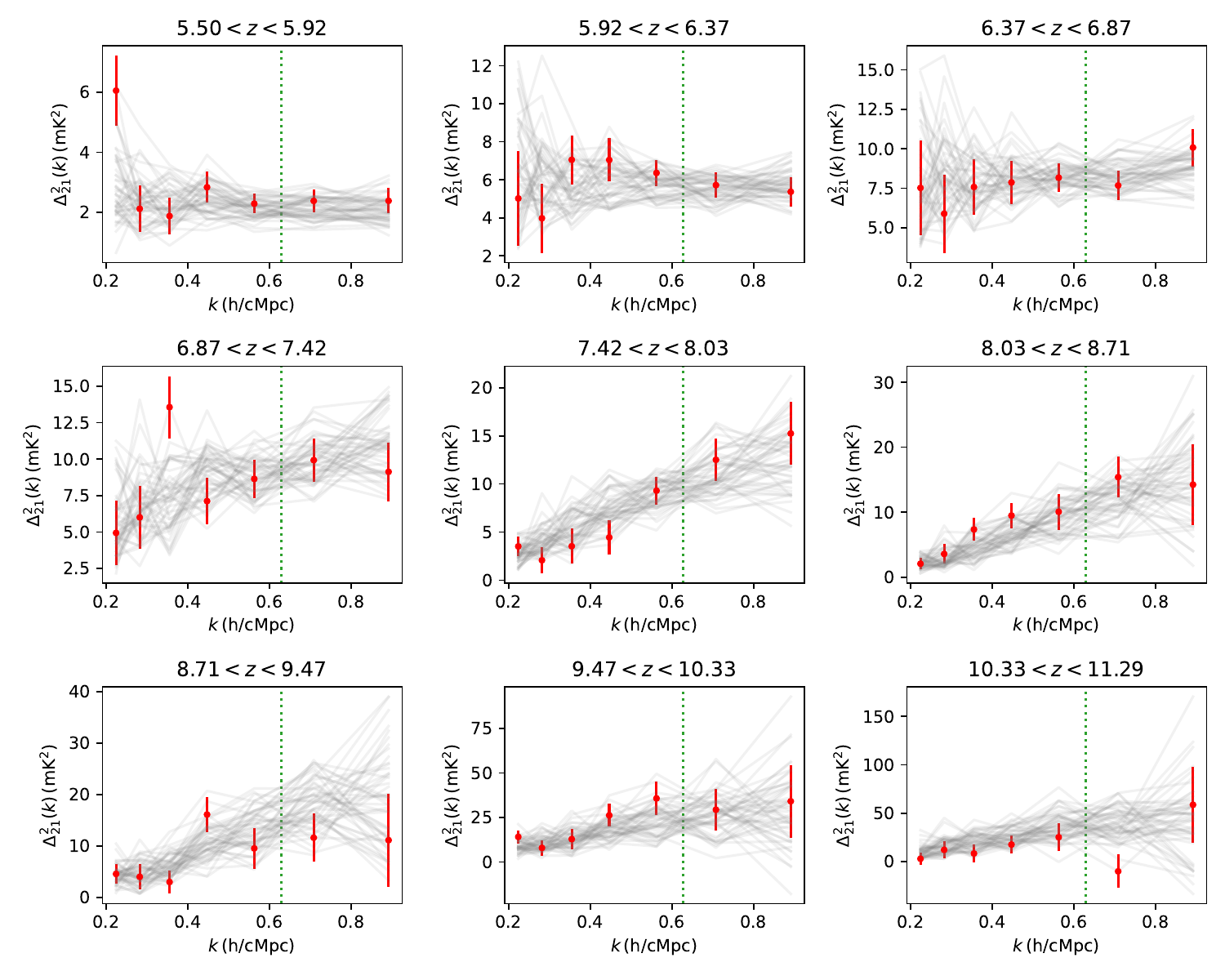}
\caption{Mock power spectra at different frequency bands corresponding to the redshift ranges mentioned at the top of individual panels. The gray lines show power spectra, including both the cosmological signal and telescope noise, for 50 individual realizations that have been used for computing the covariance. The red points with error-bars are the mock data points used for the analysis in the paper. These data points essentially correspond to the power spectrum for one of the randomly chosen realizations. The error-bars on the data points represent the diagonal terms of the covariance matrix.  All the power spectra are computed ignoring the modes in the foreground wedge. The green dotted vertical lines denote the largest $k$ used for the coarse resolution simulation.}
\label{fig:mock_powspec_data}
\end{figure*}

The mock power spectra at different redshift bins are shown in \fig{fig:mock_powspec_data}. The errorbars on the data points represent standard deviation $\sqrt{\Sigma^\mathrm{data}_{ii}}$ obtained from the diagonal terms of the covariance matrix. In addition to the mock data with error-bars, we also show the power spectra obtained from different realizations using gray lines. The green dotted vertical lines denote the largest $k$-mode used for the analysis with the coarse resolution simulation. One may notice significant jumps in certain data points. These fluctuations are due to the telescope noise added to the cosmological signal, with the magnitude and location of these jumps contingent upon the specific noise realization selected.

\subsection{MCMC analysis using the mock data}

Let us discuss the parameter constraints obtained by comparing the theoretical models with the mock data using a conventional MCMC algorithm, which will help in understanding the typical computational resources required for such an analysis. The sampling of the parameter space and computation of the posterior distributions are carried out using the publicly available package \texttt{cobaya}\footnote{\url{https://cobaya.readthedocs.io/en/latest/}} \cite{2019ascl.soft10019T,2021JCAP...05..057T}. We choose to work with the Metropolis-Hastings sampler \cite{1953JChPh..21.1087M}, appropriate for simple uni-modal or weakly multi-modal distributions, for the analysis in this paper.

We assume the likelihood to have an explicit Gaussian form. Ideally, this would require us to compute the true theoretical power spectrum for each parameter set, which would be the average of a large number of realizations of the initial cosmological field. However, computing the ionization field for a large number of redshifts for multiple realizations while spanning the parameter space would be unrealistically expensive for our MCMC analysis. Hence we take only a particular realization of the initial density field, different from the ones used to compute the mock data, to compute the theoretical power spectra. In these default MCMC runs, the theoretical model predictions are computed with the same specifications as used for the mock data, i.e., a box of $128 \hcMpc$ with $64^3$ grid cells. We shall refer to these MCMC runs as ``Full'' runs.

To account for the inherent stochasticity in the simulated power spectra because of using a particular realization, we add the cosmic variance of the simulation $\Sigma^\mathrm{sim}_{ij}$ to the covariance matrix computed in the previous section
\be
\Sigma^\mathrm{tot}_{ij} = \Sigma^\mathrm{data}_{ij} + \Sigma^\mathrm{sim}_{ij}.
\label{eq:Sigma_tot}
\ee
While computing $\Sigma^\mathrm{sim}_{ij}$ from the simulations, we do \emph{not} add noise but do include the foreground wedge effects to account for any possible bias arising by discarding particular modes. Note that the cosmic variance contained in $\Sigma^\mathrm{sim}_{ij}$ can, in principle, be different from that in $\Sigma^\mathrm{data}_{ij}$, e.g., when the simulation specifications used for the theoretical predictions are different from the ones used for constructing the mock data. Adding this extra contribution would degrade the constraints on the parameter posteriors, however, this allows us to obtain unbiased constraints on the parameters without compromising on the computational efficiency. While running the MCMC, we make another simplifying assumption that the covariance $\Sigma^\mathrm{sim}_{ij}$ is computed only for the fiducial parameter values (i.e., the ones used for generating the mock data). 

\begin{figure*}
\centering
\includegraphics[width=\textwidth]{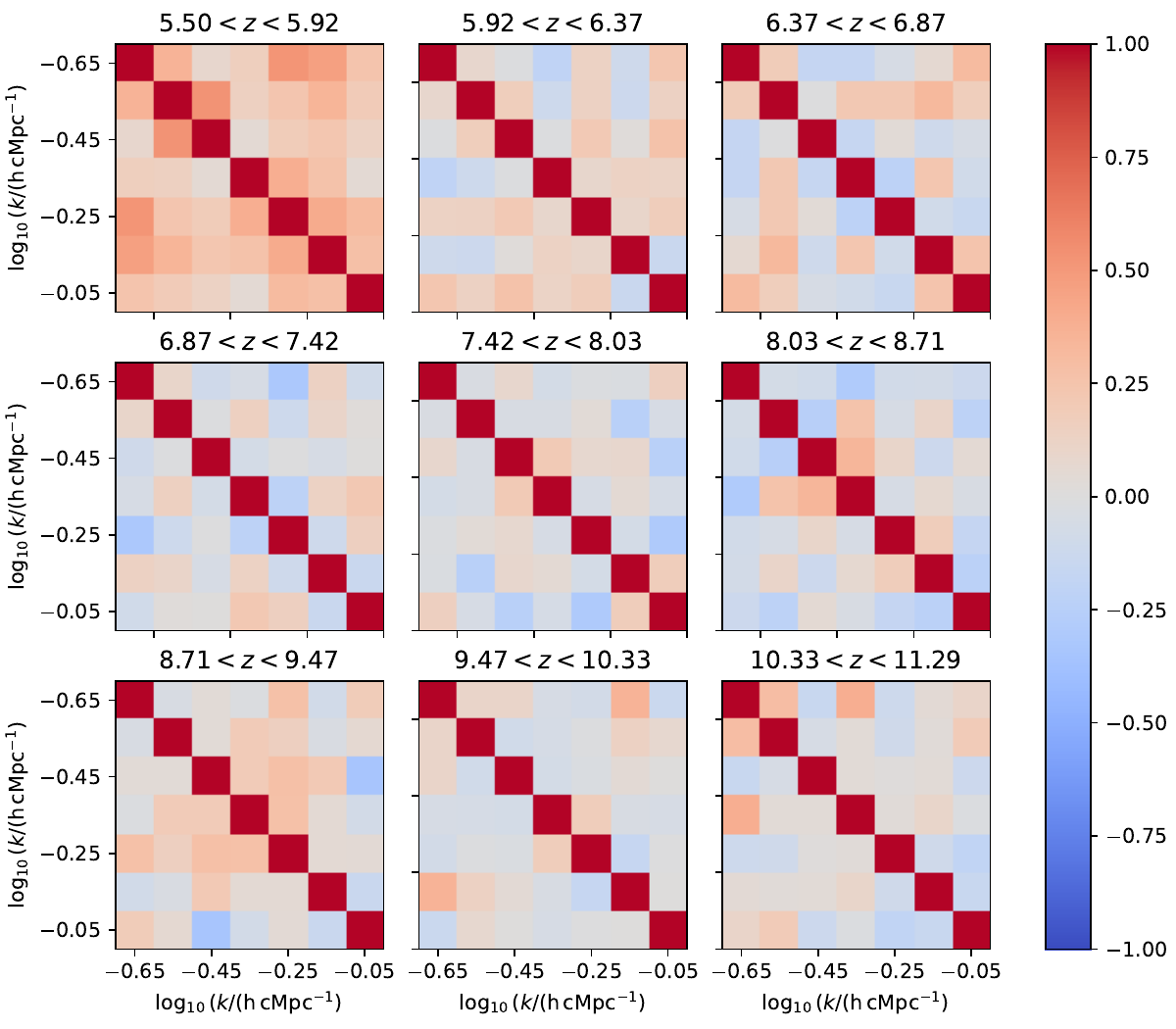}
\caption{The correlation matrix corresponding to the covariance $\Sigma^\mathrm{tot}_{ij}$ for different frequency bands, as indicated by their respective redshift ranges mentioned at the top of each panel. These matrices are obtained using 50 realizations of the cosmological field and telescope noise.}
\label{fig:correlation_matrix}
\end{figure*}

We show the correlation matrix corresponding to $\Sigma^\mathrm{tot}_{ij}$ in \fig{fig:correlation_matrix} for the 9 frequency bands used in this work. As can be seen, the off-diagonal correlations between different $k$-bins can be significant. In particular, there are pronounced positive correlations at lower redshifts (upper left panel). It is worth noting that our findings are qualitatively similar to other studies \cite{2017MNRAS.464.2992M, 2019MNRAS.487.4951S, 2023MNRAS.524.4239P}, although the specifics of computing the covariance vary across these different works.

The likelihood for our analysis is then given by
\be
\mathcal{L}\left(\Delta^{2, \mathrm{data}}_{21} \bigm | \vec{\theta}\right) = \exp\left[-\f{1}{2} \chi^2\left(\Delta^{2, \mathrm{data}}_{21} \bigm | \vec{\theta}\right)\right],
\ee
where
\be
\chi^2\left(\Delta^{2, \mathrm{data}}_{21} \bigm | \vec{\theta}\right) = \sum_{i, j = 1}^{N_\mathrm{data}} 
\left[\Delta^{2, \mathrm{sim}}_{21}(k_i; \vec{\theta}) - \Delta^{2, \mathrm{data}}_{21}(k_i) \right] 
\quad \Sigma^\mathrm{tot}_{ij} \quad
\left[\Delta^{2, \mathrm{sim}}_{21}(k_j; \vec{\theta}) - \Delta^{2, \mathrm{data}}_{21}(k_j) \right].
\label{eq:likelihood}
\ee
In the above expression, $\Delta^{2, \mathrm{data}}_{21}(k_i)$ is the mock data at the $i$th point and $\Delta^{2, \mathrm{sim}}_{21}(k_i; \vec{\theta})$ is corresponding prediction of the simulation for model parameters $\vec{\theta}$.

We consider three different parametrizations of our model for the MCMC runs:
\begin{enumerate}

\item \textbf{2p:} In this case we take a two-parameter model with $\zeta_0$ and $M_{\mathrm{min}, 0}$ as the free parameters and fix $\alpha_{\zeta, z} = \alpha_{M_\mathrm{min}, z} = 0$ to their true values.

\item \textbf{3p:} In this case we allow $\alpha_{\zeta, z}$ to be free in addition to $\zeta_0$ and $M_{\mathrm{min}, 0}$ but keep $\alpha_{M_\mathrm{min}, z} = 0$ fixed to the true value.

\item \textbf{4p:} In this case all the four parameters are allowed to vary.

\end{enumerate}

Although \textbf{4p} is the case which is of most interest to us because of its generality, we still study the other two cases in some detail to understand how our method scales with the dimensionality of the parameter space. Further, the \textbf{2p} case, because of its low dimensionality, allows us to experiment with several features of the emulator which become difficult when the dimensionality of the parameter space increases.

Instead of $\zeta_0$ and $M_{\mathrm{min}, 0}$, we choose their logarithms as the free parameters for probing larger ranges. For the free parameters, we choose sufficiently wide uniform priors, considering the possibility that their values may be highly uncertain at the outset of the MCMC analysis. The priors used are: 
\begin{itemize}
\item $\log \zeta_0 \in [0, 10]$ (all runs), 
\item $\log \left(M_{\mathrm{min}, 0} / \Msun\right) \in [7, 11]$ (all runs), 
\item $\alpha_{\zeta, z} \in [-3, 3]$ (\textbf{3p} \& \textbf{4p}),
\item $\alpha_{M_\mathrm{min}, z} \in [-8, 8]$ (only \textbf{4p}).
\end{itemize}
All the runs are assumed to be converged when the Gelman-Rubin statistic $R - 1 < 0.01$ \citet{1992StaSc...7..457G}. We discard the first $30\%$ of the samples as burn-in and consider only the rest for further analysis.

\begin{figure}
\centering
\includegraphics[width=0.65\columnwidth]{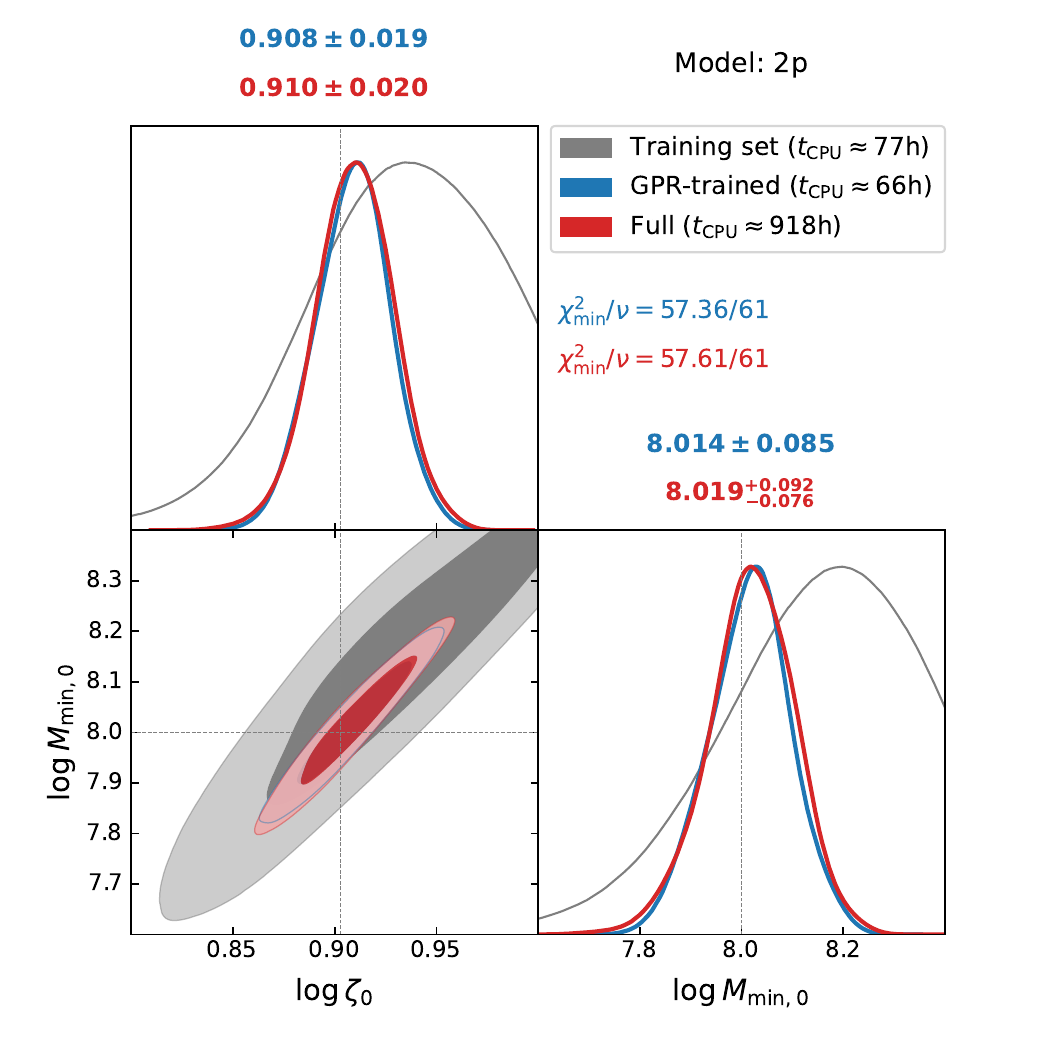}
\caption{Parameter posterior distributions for the \textbf{2p} parametrization of the model. The diagonal panels show the one-dimensional distribution, while the contour plot represents the two-dimensional joint distribution. The contour levels represent 68\% and 95\% confidence levels. The mean and 68\% confidence intervals are denoted above the one-dimensional posterior of the respective parameters. 
The gray colour plots, denoted by ``Training set'', represent the MCMC run with coarse resolution simulations, the corresponding posteriors being used to draw the parameter values for the training set for the GPR emulator. The blue colour represents the results obtained using the ``GPR-trained'' emulator. The red colour plots are for the ``Full'' MCMC run.
The core CPU hours $t_\mathrm{CPU}$ are mentioned for each case in the legend. Note that for the ``GPR-trained'' case, $t_\mathrm{CPU}$ is essentially the time taken for generating the $\chi^2$ for the training samples using the default resolution simulations, the actual training and MCMC take negligible time. The total time taken for obtaining the posteriors starting from scratch would be the sum of $t_\mathrm{CPU}$ for the ``Training set'' and ``GPR-trained'' case, which is significantly smaller than the $t_\mathrm{CPU}$ for the ``Full'' run.
We have also mentioned the best-fit $\chi^2$ per degree of freedom $\nu$ obtained using the emulator (blue) and the ``Full'' run (red).}
\label{fig:gpr_2p_hires_lores}
\end{figure}

\begin{figure*}
\centering
\includegraphics[width=0.65\textwidth]{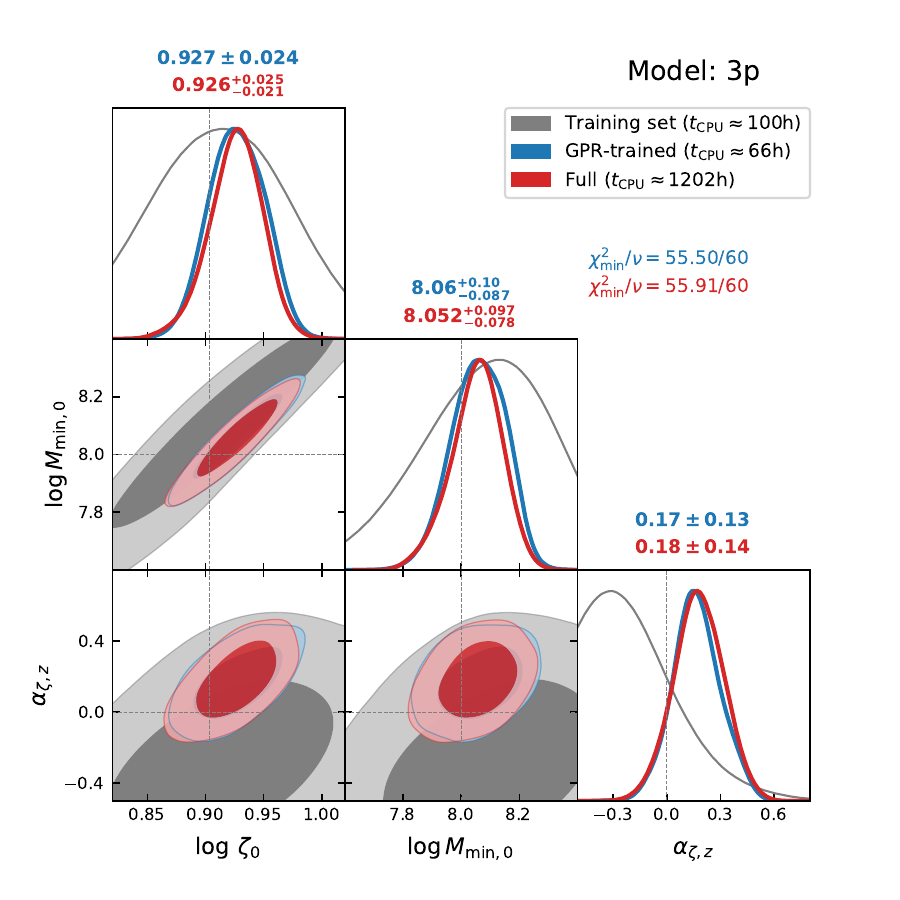}
\caption{Same as \fig{fig:gpr_2p_hires_lores} but for the \textbf{3p} parametrization.}
\label{fig:gpr_3pz_hires_lores}
\end{figure*}

\begin{figure*}
\centering
\includegraphics[width=0.9\textwidth]{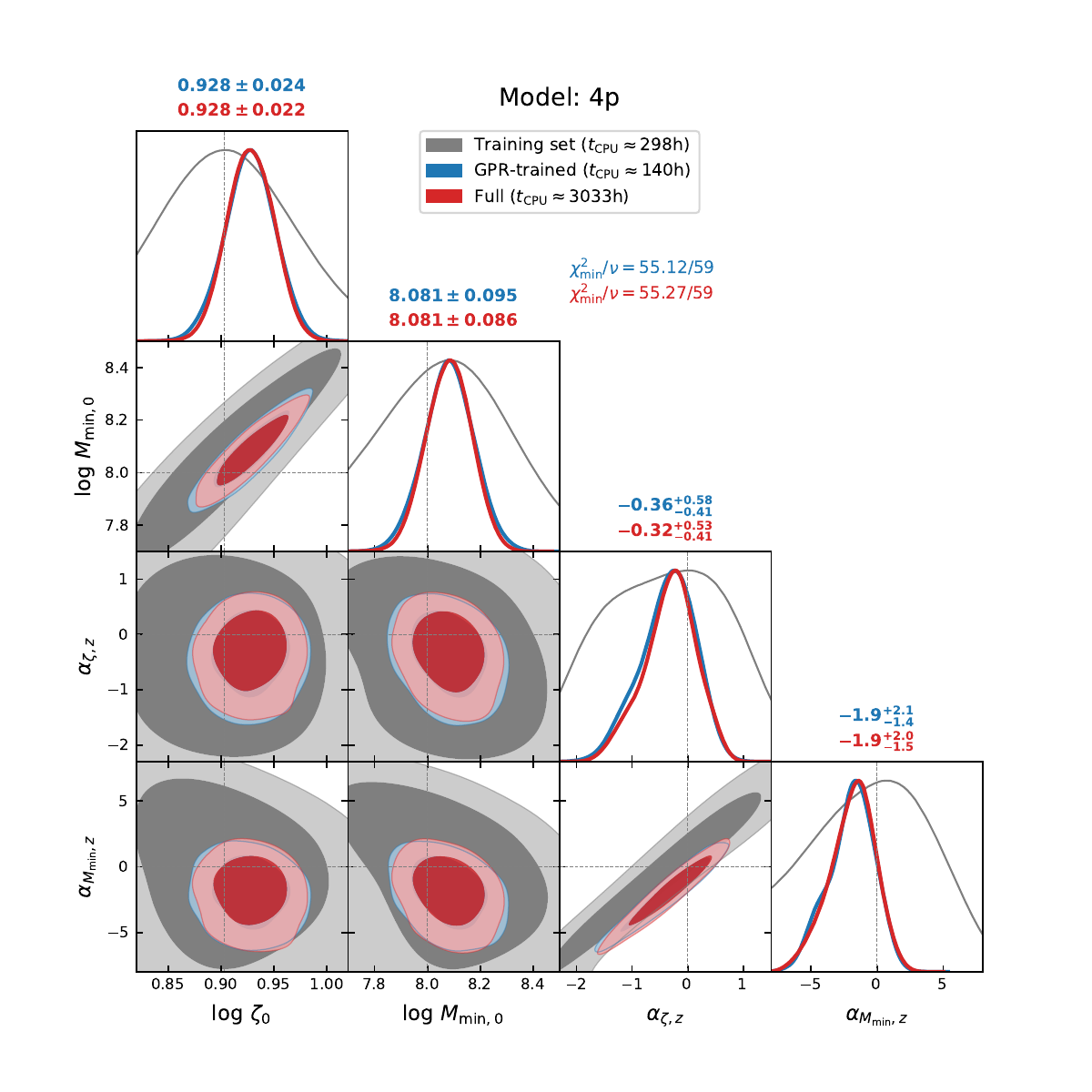}
\caption{Same as \fig{fig:gpr_2p_hires_lores} but for the \textbf{4p} parametrization.}
\label{fig:gpr_4p_hires_lores}
\end{figure*}

The results of the ``Full'' MCMC runs for the three cases are shown by red lines and contours in \figsthree{fig:gpr_2p_hires_lores}{fig:gpr_3pz_hires_lores}{fig:gpr_4p_hires_lores}, respectively. The input values of the parameters used for creating the mock data are shown by dashed lines. It is clear that the input values are always within the 1-$\sigma$ confidence contours and thus the MCMC provides unbiased estimates of the free parameters for all the three cases. To judge how good the fits are, the $\chi^2$ values for the respective best-fit models $\chi^2_\mathrm{min}$ per degree of freedom $\nu$ are also mentioned in the respective figures in red. We can see $\chi^2_\mathrm{min} / \nu \lesssim 1$ for all the cases, thus indicating that our estimates of the covariance are realistic.

In the figure legend, we also mention the total CPU time or core hours, denoted as $t_\mathrm{CPU}$, required for these MCMC runs to finish. To understand these numbers, note that the \textbf{2p} run required $\sim 1.3 \times 10^4$ evaluations of the likelihood, each consuming $\sim 4$~minutes. The corresponding number of evaluations for the \textbf{3p} and \textbf{4p} runs were $\sim 1.6 \times 10^4$ and $\sim 5 \times 10^4$, respectively. The \textbf{4p} run involved a significantly higher number of evaluations compared to the other two. This difference could be because of higher dimensionality of the parameter space, combined with the fact that $\alpha_{M_{\mathrm{min}}, z}$ is relatively poorly constrained, requiring more steps to achieve convergence. At the same time, it must be noted that the number of steps required for MCMC runs can fluctuate by $\sim 30\%$ due to the inherent stochasticity in choosing the subsequent steps while spanning the parameter space, as we checked by completing a few additional runs with the \textbf{2p} model. In the case of the \textbf{4p} run, the CPU time $t_\mathrm{CPU} \approx 3033$~hours translates to $\sim$ a month to complete the run using 4 processors on a standard desktop. The run could possibly be completed in a shorter time using more processors, however, the scaling of the Metropolis-Hastings algorithm with respect to the number of processes is not linear in our case, particularly when the number of chains exceeds 8 (again, based on our experience with the \textbf{2p} model).

\section{GPR for SCRIPT}
\label{sec:GPR}

It is clear from the preceding discussions that the conventional MCMC runs require a large number of evaluations of the likelihood, and hence substantial computing time to converge satisfactorily. Consequently, there remains ample scope for exploring reliable ways to compute the posterior while minimizing the number of calls to the simulation, possible to achieve through the usage of an emulator.

\subsection{Features of the GPR-based emulator}
\label{subsec:gpr_emulator}

We introduced a GPR-based likelihood emulator for our semi-numerical code \texttt{SCRIPT} in \citetalias{2023MNRAS.526.3920M}, the one employed in this study follows the same steps. For completeness, let us summarize the main points of the emulator here:

\begin{itemize}

\item Our aim is to interpolate the log-likelihood function $\chi^2\left(\Delta^{2, \mathrm{data}}_{21} \bigm | \vec{\theta}\right)$ for any $\vec{\theta}$, given a training set $\chi^2\left(\Delta^{2, \mathrm{data}}_{21} \bigm | \vec{\theta}_t\right)$, even when $\vec{\theta}_t$ are sparsely sampled. Unlike emulators that interpolate observables \cite{2017ApJ...848...23K,2017MNRAS.468.3869S,2018MNRAS.475.1213S,2019MNRAS.483.2907J,2020MNRAS.493.4728G,2022MNRAS.514L..31M,2023arXiv230715577L,2023MNRAS.524.4239P,2023arXiv230905697B}, our approach focuses on interpolating a single function. The advantage of our approach is that it avoids potential complications arising from features inherent in the different observables. The caveat is that the method depends on the specific data set in use (in our case $\Delta^{2, \mathrm{data}}_{21}$) used, hence the emulator needs to be retrained when the data set is different. This limitation can be mitigated to some extent if one is able to store the observables from the simulations.

\item We employ GPR for interpolation. As is well known, GPR is a non-parametric Bayesian technique that, given a set of data points, offers a means to predict the function at arbitrary values \cite{2006gpml.book.....R}. It essentially involves a set of random variables with a joint Gaussian distribution, determined by a kernel that specifies the covariance of the variables. The kernel typically includes several free parameters, known as hyperparameters, which are optimized by comparing the GPR-predicted function with the actual values from the training set. In our case, we opt for the Matern kernel with an order of $\nu = 3/2$, which has proven to be adequate.

\item The hyperparameters are constrained through an iterative process. It starts with a small fraction of the training samples, and aims to maximize the log-marginal likelihood, essentially a measure of the difference between the predicted and actual values of the function to be interpolated. We use the Anisotropic Simulated Annealing algorithm for this maximization \cite{2022arXiv220507906P}, although other robust multi-dimensional minimization algorithms can also be used. At each iteration, we cross-validate the optimized hyperparameters by comparing the predicted $\chi^2$ with a subset of training samples that were \emph{not} used during the optimization. If the cross-validation is not successful, we increase the training sample size by 10\% and repeat the optimization. This iteration continues until the cross-validation succeeds.

\item The success of the cross-validation is determined by the threshold parameter \texttt{cv\_thresh}. If the 16 and 84 percentiles of the relative differences between the predicted and true $\chi^2$ fall below this threshold, the training is assumed to be converged.

\item Once the training is converged, the values of the hyperparameters along with the specific subset of the training set that led to convergence are stored for the further interpolation, effectively forming the emulator.  Storing the hyperparameter values as well as training subset makes the emulator fully portable. We emphasize that, since we only emulate a single function (the log-likelihood), the storage requirements are minimal. In case the training does not converge, it indicates that either the training sample size needs to be increased or the convergence criterion should be relaxed, e.g., by increasing the value of \texttt{cv\_thresh}. We will discuss the method of addressing this in \secn{subsec:contruct_emuletor}.

\end{itemize}

\subsection{Constructing the training set}
\label{subsec:construct_training}

As discussed in \citetalias{2023MNRAS.526.3920M}, a crucial step in constructing the emulator is to select an appropriate training set, subsets of which are used in the GPR training. We require two conditions to be satisfied while choosing the parameter values $\vec{\theta}_t$ of the training set, namely, (i) the training should converge for a reasonably small \texttt{cv\_thresh} using a minimum number of samples, and (ii) the resulting emulator should faithfully reproduce the posterior distributions, in particular, the 1-$\sigma$ and 2-$\sigma$ confidence regions. This immediately indicates that the samples should be concentrated in regions of high posterior probability, so that errors accrued by the likelihood emulator when subsequently exploring this region in an MCMC run are kept at a minimum. Finding these regions can be a challenging task if the priors are considerably wider than the eventual posterior (which is the case for our analysis).\footnote{We explored the feasibility of constructing a training set using conventional sampling methods, such as the Latin hypercube, for the simplest parametrization \textbf{2p}. It turned out that, for the broad priors considered in this paper, the resulting emulator failed to generate any posteriors consistent with the ``Full'' MCMC run. This discrepancy persisted even with a sample size as large as $\sim 10^4$, similar to the number of samples needed for the ``Full'' run.} This challenge can be overcome, as shown in \citetalias{2023MNRAS.526.3920M}, by running an MCMC using a coarse resolution simulation setup and sampling the resulting posterior to create the training set. \emph{This method is expected to work because the 21~cm power spectra computed using \texttt{SCRIPT} converges with respect to resolution,} as shown in our earlier works \citep{2018MNRAS.481.3821C,2023MNRAS.521.4140M}.

We thus consider a setup with grid cells of size $4 \hcMpc$ which leaves us with only $32^3$ grid cells. The corresponding simulations are $\sim 8-10$ times faster than the default case. The downside of using a coarse resolution simulation is that the largest $k$-modes in the mock data are not accessible, hence we need to discard 2 points per redshift bin. This is not a major issue because the simulations that will be used to actually evaluate the likelihood and build the emulator later will have a higher resolution that can account for these large-$k$ modes. The largest $k$-mode reliably probed by this coarse resolution simulation is indicated by the green dashed vertical lines in \fig{fig:mock_powspec_data}.

Using a simulation of a different resolution implies that the cosmic variance $\Sigma^\mathrm{sim}_{ij}$ of the simulation would be different than the default resolution. This, of course, is straightforward to implement using \eqn{eq:Sigma_tot}. There is another, perhaps more important effect, which needs to be accounted for. Since the mock data does not remain the same for the two resolutions, it is possible that the resultant best-fit and the posterior contours obtained from the coarse resolution simulations are shifted compared to those in the default resolution. In that case, the regions sampled by the coarse resolution simulations, and hence the training set, might not enclose sufficient volumes around the high-probability regions of the default resolution run. This can lead to inaccurate interpolation while running the MCMC using the trained GPR and hence resulting in posteriors that are biased. Similar cases, in the context of observations other than 21~cm, were studied in \citetalias{2023MNRAS.526.3920M} and it was found that the posteriors from which the training samples were drawn must enclose the 2-$\sigma$ contours of the true posteriors to ensure robust interpolation.

Since the true posterior distribution is not known beforehand, one needs to come up with a working solution to ensure that the training set is drawn from a reliable distribution of parameters. The solution we introduce in this paper is to simply \emph{increase the effective errors} by scaling the covariance $\Sigma^\mathrm{sim}_{ij}$ of the simulation by a factor larger than unity so that the posteriors are wider. We thus calculate the total covariance as
\be
\Sigma^\mathrm{tot, scaled}_{ij} = \Sigma^\mathrm{data}_{ij} + \lambda \Sigma^\mathrm{sim}_{ij},~~\lambda > 1.
\label{eq:Sigma_scaled}
\ee
The choice of $\lambda$ needs some intelligent guesswork. If $\lambda$ is too small, we run the risk of the training samples not covering the relevant regions of the true posterior distribution. On the other hand, too large a $\lambda$ may lead to sparse sampling of the high-probability region and consequently require a larger training set. Assuming that the best-fits (or the means) of the posterior distributions obtained from the two resolutions do not deviate by more than 3-$\sigma$ of each other, one optimal choice could be $\lambda \sim 9$. However, it is still possible that the posterior contours have highly non-elliptical and/or highly correlated shapes, in which case one may be inclined to consider larger $\lambda$ values. We did carry out some tests for the \textbf{2p} model and found that $8 \leq \lambda \leq 16$ provides decent results for our case. We hence choose the default value $\lambda = 8$ for our analysis, but will discuss deviations from this in \secn{sec:diff_lambda}. Because of this arbitrary scaling, the posteriors obtained using the coarse resolution simulations may not contain any useful information on the eventual parameter constraints. This, however, is of no real concern as these posteriors are used only for constructing the training set.

With the above modifications, we run the MCMC for the three parametrizations using the coarse resolution simulation. These runs complete in considerably less time than those with the default resolutions. The posterior distributions from which the training samples are drawn are shown in gray in \figsthree{fig:gpr_2p_hires_lores}{fig:gpr_3pz_hires_lores}{fig:gpr_4p_hires_lores} for the three respective parametrizations. Comparing the distributions with the ``Full'' MCMC (red), we can clearly see that they are considerably wider for these runs. This broadening of the distribution is mostly contributed by the scaling factor $\lambda$, with some small contribution arising because the number of data points used in the coarse simulations is slightly less than the ``Full'' run. We also mentioned in the legends, as ``Training Set'', the computing times required for these runs to complete. As one can see, the runs take almost an order of magntiude less time compared to the default resolution ones.

Given the posteriors, and for a given choice of the number of training samples, we sample the posterior distribution randomly (discarding the first $30\%$ samples as burn-in) to obtain $\vec{\theta}_t$ and compute the $\chi^2\left(\Delta^{2, \mathrm{data}}_{21} \bigm | \vec{\theta}_t\right)$ using the \emph{default} resolution (i.e., $64^3$ grid points). The requirement of the default resolution for this calculation can make this step relatively expensive if the number of training samples become too large. Hence it is critical to keep this number as small as possible, and we will discuss how we achieve that in the next section. The $\chi^2\left(\Delta^{2, \mathrm{data}}_{21} \bigm | \vec{\theta}_t\right)$ thus constructed are then used for training the emulator using the iterative method outlined earlier in \secn{subsec:gpr_emulator}.

\subsection{Constructing the emulator}
\label{subsec:contruct_emuletor}

To fix the optimal size of the training set, we start with $\sim 1000$ samples drawn from the posterior distribution, which is obtained using the coarse resolution simulations. We calculate $\chi^2\left(\Delta^{2, \mathrm{data}}_{21} \bigm | \vec{\theta}_t\right)$ for this parameter set $\vec{\theta}_t$ using the \emph{default} resolution (i.e., $64^3$ grid points). These samples $\vec{\theta}_t$ and the corresponding $\chi^2\left(\Delta^{2, \mathrm{data}}_{21} \bigm | \vec{\theta}_t\right)$ are then passed on to the GPR training process, where we attempt to train the emulator with a stringent $\mathtt{cv\_thresh} = 0.01$. Note that the training process is inherently iterative, see \secn{subsec:gpr_emulator}, and may not necessarily utilize all the samples $\vec{\theta}_t$ supplied. In case we do not achieve successful convergence with this training set, we increase $\mathtt{cv\_thresh}$ gradually in steps of $0.01$ until the training converges. If no convergence is achieved till $\mathtt{cv\_thresh} = 0.05$, we increase the training samples by $50\%$ by selecting new samples from the posterior distribution. It is important to keep in mind that this increase in the sample size involves additional likelihood evaluations using the default resolution, and thus may add considerably to the time taken for constructing the emulator. With this increased sample size, we reset $\mathtt{cv\_thresh} = 0.01$ and repeat the previous steps until the training converges.

The eventual sample size required, which is equal to the number of calls to the default resolution simulations, for the three cases \textbf{2p}, \textbf{3p} and \textbf{4p} turn out to be $936$, $944$ and $2006$, respectively. We achieve $\mathtt{cv\_thresh} = 0.01$ in all the three cases. Note that the number of samples required to train for the \textbf{4p} case is twice the other two, possibly because of the higher dimensionality. While a significant time is required for computing the likelihoods for the training set, the training itself typically takes only $\sim$ a minute to about a few minutes on a single processor.

After constructing the converged cross-validated emulator, we use it to run the MCMC. These runs are extremely fast and each completes in about a minute using only one processor, hence this contribution too can be ignored while computing the total time taken for the emulator-based analysis.

\subsection{Posterior distributions using the emulator}

The results for the three parametrizations obtained using the emulator are shown in \figsthree{fig:gpr_2p_hires_lores}{fig:gpr_3pz_hires_lores}{fig:gpr_4p_hires_lores} in blue. \emph{The key takeaway is the remarkable agreement between the posteriors derived from the emulator and those obtained through the ``Full'' MCMC run (red).} Furthermore, the mean values and 1-$\sigma$ limits for the parameters, mentioned above the panels showing their respective one-dimensional posteriors, agree between the two methods. The differences in the mean values are all significantly smaller than the 1-$\sigma$ uncertainties, and the 1-$\sigma$ uncertainties themselves between the two methods agree to within $\lesssim 10\%$. Moreover, the minimum $\chi^2_\mathrm{min}$ obtained from both methods matches consistently across all three models. Overall, regardless of the dimensionality of the problem, it is evident that the emulator consistently provides parameter posteriors that closely match with the true posterior distributions.

We now turn our attention to the computational time required to obtain the posteriors using the emulator in comparison to the ``Full'' MCMC run. Recall that the computing time for the emulator has two contributions:

\begin{enumerate}

\item[(i)] The first involves the MCMC run using the coarser resolution simulations, the posteriors of which are used to generate the parameter sets $\vec{\theta}_t$ for the training samples, as discussed in \secn{subsec:construct_training}. We denote the CPU time for this step as ``Training Set'' in the legends.

\item[(ii)] The second part is related to the likelihood evaluations using the default resolution simulation for the parameters corresponding to the chosen training set. Given that the default resolution simulations can be relatively computationally demanding, this step often consumes a significant amount of time. The corresponding CPU time is indicated as ``GPR-trained'' in the legends. 

It is worth noting that once these two steps are completed, both the training and subsequent MCMC runs require negligible time (approximately a few minutes) and can therefore be ignored in this context.

\end{enumerate}

For all the three parametrizations, we find that the total time required to obtain the posteriors is significantly less when we use the GPR emulator, typically by a factor of $\sim 7$. This is almost an order of magnitude reduction in computing time and could potentially be an extremely useful method for constraining the parameters. The fact that the gain in speed remains almost independent of the dimensionality of the problem (to the extent of models studied here) is also promising, suggesting that we can possibly extend the approach to more complex parametrizations.

One point to be emphasized here is that, when computing the CPU time for the emulator, we take into account the time required for both generating the training samples and carrying out the training itself. In other words, the time taken for the emulator would be that where one starts from scratch without access to any previously run models. Our method thus proves particularly useful when one wants to explore new parametrizations of the model, where the use of previously generated training samples may not be feasible.

In the next couple of sections, we study a few variants of the default run. We concentrate only on the \textbf{4p} case for these variants.

\subsection{Choosing a different $\lambda$}
\label{sec:diff_lambda}

\begin{figure*}
\includegraphics[width=0.9\textwidth]{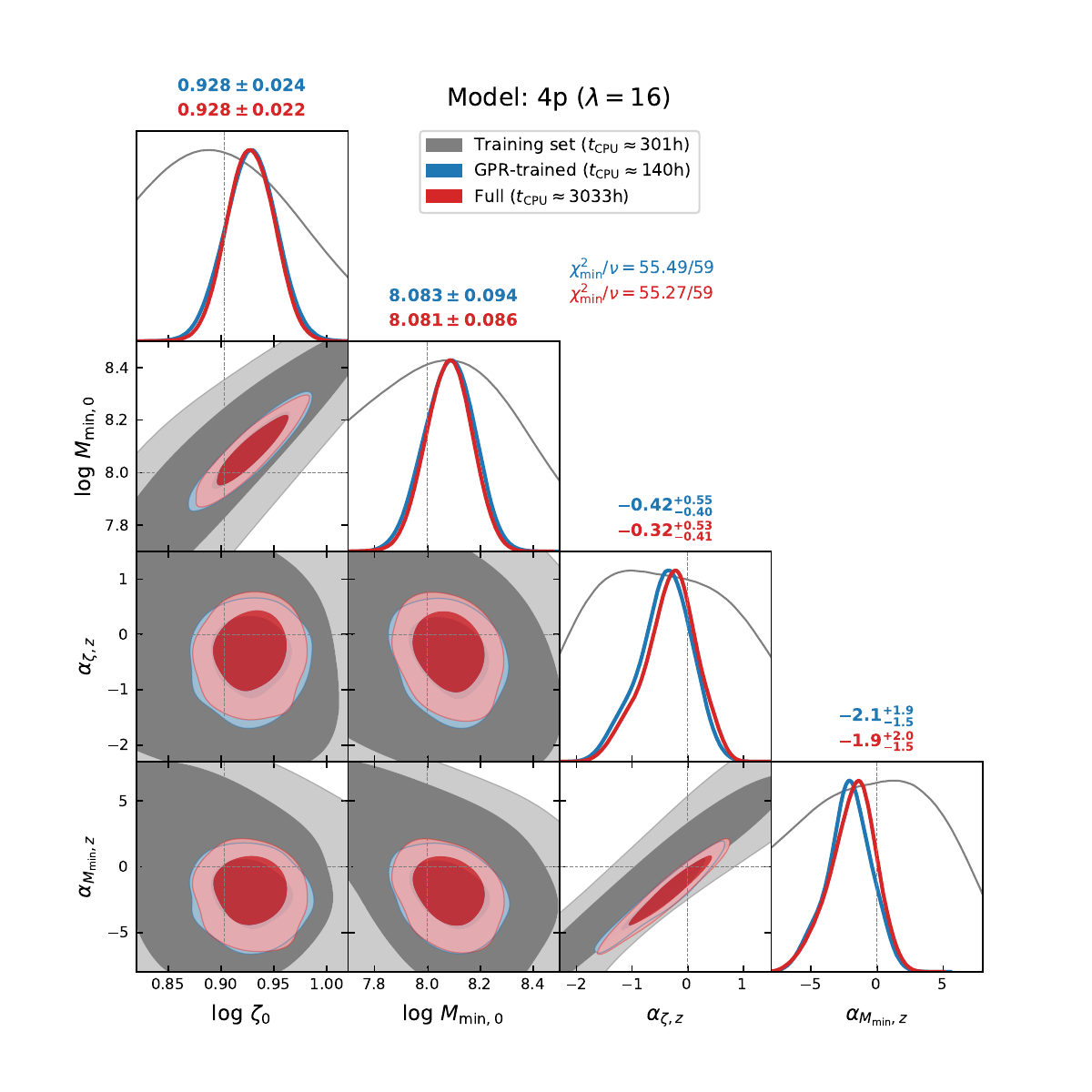}
\caption{Same as \fig{fig:gpr_4p_hires_lores} but for $\lambda = 16$ instead of the default $\lambda=8$, see \eqn{eq:Sigma_scaled} and the related discussion for more details.}
\label{fig:gpr_4p_lam16_hires_lores}
\end{figure*}

Let us take a closer look at the two-dimensional posterior distributions for the \textbf{4p} case, as shown in \fig{fig:gpr_4p_hires_lores}. In the $\alpha_{\zeta, z} - \alpha_{M_\mathrm{min}, z}$ plane, a visual inspection reveals that the contour edges obtained from the emulator (blue contours) are quite close to the edges of the training samples (gray contours) at the bottom right. This raises the question of whether the training samples had sufficient coverage close to the edges of the 2-$\sigma$ contours and whether one can trust the training. While we already know from the ``Full'' run (red contours) that there is no cause for concern, as the posteriors obtained using the emulator match the true results, the real test comes when using the emulator for a general case without access to the ``Full'' run. In such cases, it might be prudent to verify if the results remain consistent when the coarse resolution posteriors are generated using a larger scaling factor $\lambda$, so that the range of the parameter space used for drawing the training samples widen and the samples cover the eventual posteriors reliably.

We show the results for $\lambda = 16$ in \fig{fig:gpr_4p_lam16_hires_lores}. Clearly, due to the increased value of $\lambda$, the posterior distributions from which the training samples are drawn have widened, as observed by comparing the gray regions in \figs{fig:gpr_4p_hires_lores}{fig:gpr_4p_lam16_hires_lores}. In particular, the gray posteriors in the $\alpha_{\zeta, z} - \alpha_{M_\mathrm{min}, z}$ plane for the $\lambda = 16$ case (\fig{fig:gpr_4p_lam16_hires_lores}) entirely enclose the 2-$\sigma$ contours of the ``Full'' MCMC run. The posteriors obtained using the training and subsequent GPR training, however, remain unaffected when compared to the default $\lambda = 8$ case. If anything, the agreement between the ``Full'' and GPR-based methods for the parameters $\alpha_{\zeta, z}$ and $\alpha_{M_\mathrm{min}, z}$ has slightly improved compared to the default case. Moreover, the time taken for identifying the training set and then building the emulator is only marginally ($\lesssim0.7\%$) larger than in the default case. This exercise allows us to conclude that our results as well as computational time are insensitive to the value of the scaling factor $\lambda$ as long as it is within the range $8 - 16$. Based on this study and also lessons learned in \citetalias{2023MNRAS.526.3920M}, we suggest the following approach: one initiates the process by generating the training set with $\lambda \approx 8$ and verify if the edges of the posteriors obtained using the emulator are close to the edges of the corresponding training sample distribution. In case they are not, it provides confidence in the robustness of the results. Alternatively, if the edges are close to each other, one may consider validating the posteriors by running the emulator with a larger $\lambda$, albeit at the cost of increased CPU time.

\subsection{Faster ways of generating the training samples}

\begin{figure*}
\includegraphics[width=0.9\textwidth]{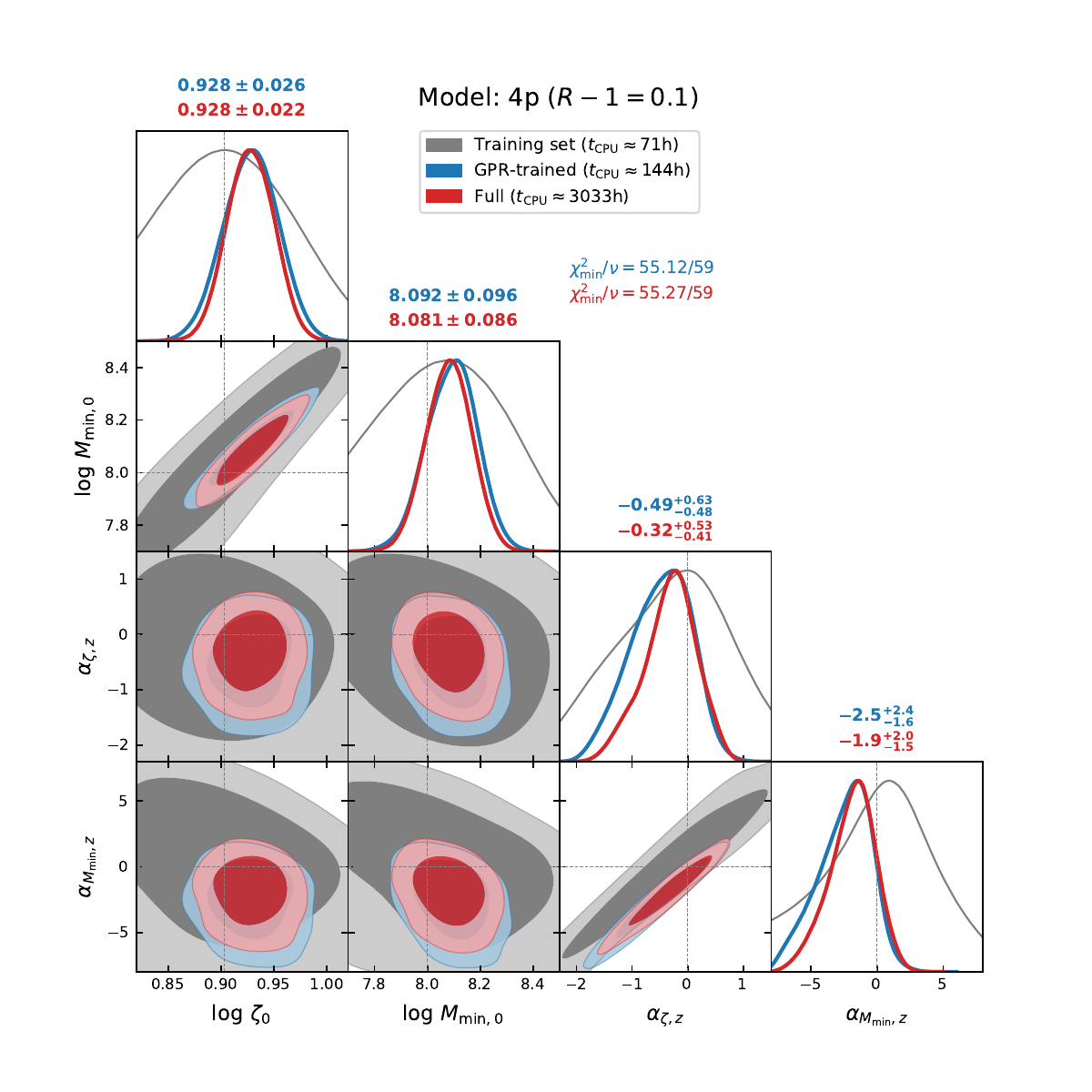}
\caption{Same as \fig{fig:gpr_4p_hires_lores} but for a less stringent convergence criterion while running the MCMC for coarse resolution simulations. The default case uses a Gelman-Rubin condition $R - 1 < 0.01$, while this uses $R - 1 < 0.1$. Recall that the posterior for these runs are used only for drawing the parameters for the training set.}
\label{fig:gpr_4p_low_Rm1_hires_lores}
\end{figure*}

For the emulator, it is clear that a large fraction of the computing time is spent in generating the posteriors that are subsequently used for drawing the training samples. As these posteriors are not used for anything other than identifying the regions of high probability, it is worth exploring the possibility of terminating the MCMC run with a less stringent convergence criterion (i.e., a higher $R - 1$).  This approach will save us some computing time at the expense of somewhat less accurate posteriors. To investigate this, we check the results for a less stringent threshold $R - 1 = 0.1$ used for running the MCMC with coarse resolution simulations. It turns out that the MCMC indeed takes substantially less time, however, the number of samples required to train the emulator, which is equal to the number of calls to the default resolution simulation, remains same ($=2006$) as the \textbf{4p} case. The results are shown in \fig{fig:gpr_4p_low_Rm1_hires_lores}.

Comparing these results with those in \fig{fig:gpr_4p_hires_lores}, the first thing to note that the time required to generate the posteriors from which training samples are drawn is only $\sim 24\%$ of the corresponding time for the default case. As a result, the speedup achieved through the emulator-based MCMC is $\sim 14$-fold, an improvement by a factor $\sim 2$ compared to the default case. However, the match between the posteriors of the GPR-trained and ``Full'' MCMC is not as good as for the default case. The error-bars are overestimated for almost all the parameters in the GPR-trained case, by up to $\sim 20\%$. The mean values of the parameters, particularly $\alpha_{\zeta, z}$ and $\alpha_{M_\mathrm{min}, z}$, are also different for the GPR-trained case. At the same time, it must be noted that the differences in the mean values are considerably smaller than the 1-$\sigma$ uncertainties.

To understand the main cause for this discrepancy, we look closely at the distribution from the which the training samples were selected (the gray contours and lines). If we compare the one-dimensional posteriors between the default and this case, in \figs{fig:gpr_4p_hires_lores}{fig:gpr_4p_low_Rm1_hires_lores}, we find that they appear quite distinct for the parameters $\alpha_{\zeta, z}$ and $\alpha_{M_\mathrm{min}, z}$. For example, the weighting towards lower values of these parameters for the \textbf{4p ($R - 1 \approx 0.1$)} case is lower compared to the default \textbf{4p} case, and the discrepancies in the posteriors appear in similar regions. Given that the total number of samples used for training the GPR  remains the same in both cases, this implies that we have fewer points with lower values of these parameters. This leads to larger inaccuracies in the emulator predictions. The analysis thus highlights the importance of concentrating on regions of high probability to obtain an accurate emulator with minimum number of likelihood evaluations.

In spite of the discrepancies, we should stress that the mismatch remains within $10 - 20 \%$. Therefore, depending on the requirements of the analysis, one can benefit significantly from reduced computing time using a less strictly converged posterior for drawing the training samples. 

\section{Summary and future prospects}
\label{sec:summary}

Theoretical models of reionization typically involve a large number of unknown parameters, and their values can be constrained only by comparing with observations. Exploring this parameter space using simulations can be computationally intensive, making emulators a valuable tool for such studies. In this work, we have presented a likelihood emulator based on GPR for our semi-numerical reionization code, \texttt{SCRIPT}, which was subsequently used for parameter inference using mock 21~cm power spectrum data and Bayesian MCMC analysis. We validated the performance of the emulator-based MCMC by comparing with a full MCMC run.

A unique aspect of our methodology, introduced in a previous work \citepalias{2023MNRAS.526.3920M}, is the utilization of coarser resolution simulations, that require significantly reduced computational time, to identify high-probability regions within the parameter space. This information is used to train the emulator. The subsequent MCMC is found to provide reasonably accurate parameter posteriors while reducing computing time by approximately an order of magnitude.

Our method is particularly beneficial for scenarios where one wants to use different parametrizations of the reionization model and constrain those parameters efficiently. The method is also useful when the prior distributions on the parameters are significantly wider than the eventual posteriors. The entire training process, involving MCMC with coarse resolution simulations and setting up the training set using default resolution, can be performed in a reasonable amount of computing time. Given our limited understanding of the galaxy properties at high redshifts and with advancement in observations and theoretical studies, one may be interested in trying out different parametrizations of the reionization sources, occasionally requiring broad prior ranges. Our approach can be very effective in such situations.

While our analysis is promising, certain aspects still have scope for improvements. For example, our treatment of the covariance between the different data points remains approximate, as we do not account for correlation between different frequency bands or compute the covariance for every point in the parameter space. Our analysis can also be extended to combine other observational data sets with the 21~cm power spectra. Furthermore, our simulations have a limitation that the grid cells must have a length $\gtrsim 2 \hcMpc$ to relate the Eulerian grid densities to their Lagrangian counterparts -- an essential step for generating the sub-grid haloes. It is worth exploring the feasibility of training an emulator with small-volume high-resolution simulations and using it to predict the halo collapse fractions in a large-volume coarse grid box. Additionally, we assume an explicitly Gaussian likelihood, which can be extended through more general techniques such as Simulation Based Inference (SBI), already in use for parameter inference from 21~cm power spectrum \cite{2022ApJ...933..236Z,2023MNRAS.525.6097S,2023MNRAS.524.4239P}. These are directions we intend to explore in future work.

However, for almost all parameter inference techniques based on machine learning, it is useful to have a prior knowledge about the regions of high probability where the posteriors are likely to be concentrated. Our approach, involving a coarse resolution simulation to obtain this information with minimal computational resources, can be quite useful in this regard, particularly for situations where one wants to consider different parametrizations of the model.

\section*{Acknowledgments}
The research of AP is supported by the Associates scheme of ICTP, Trieste.

\section*{Data Availability}

The data generated during this work will be made available upon reasonable request to the corresponding author.

\bibliographystyle{JHEP}
\bibliography{21cm_GPR} 

\end{document}